\documentclass[prl,aps,twocolumn,floatfix,superscriptaddress,preprintnumbers]{revtex4-1}

\usepackage{mathrsfs}
\usepackage{amsfonts}
\usepackage{amsmath,amssymb,bm}
\usepackage{array}
\usepackage{verbatim}
\usepackage{epsfig}
\usepackage{graphicx}
\usepackage[colorlinks=true,linkcolor=blue]{hyperref}
\usepackage[normalem]{ulem}
\usepackage{xcolor}
\usepackage{ulem}
\usepackage{multirow}

\begin{document}

\preprint{JLAB-THY-19-2927, Nikhef-2019-012}

\title{Strange quark suppression from a simultaneous Monte Carlo analysis
	\\ of parton distributions and fragmentation functions}

\author{N. Sato}
\affiliation{Jefferson Lab, Newport News, VA 23606, USA}
\author{C. Andres}
\affiliation{Jefferson Lab, Newport News, VA 23606, USA}
\author{J. J. Ethier}
\affiliation{Department of Physics and Astronomy,
	Vrije Universiteit Amsterdam, 1081 HV Amsterdam,\\
        and Nikhef Theory Group, Science Park 105, 1098 XG Amsterdam,
	The Netherlands	\\
        \vspace*{0.2cm}
        {\bf Jefferson Lab Angular Momentum (JAM) Collaboration
        \vspace*{0.2cm}}}
\author{W. Melnitchouk}
\affiliation{Jefferson Lab, Newport News, VA 23606, USA}

\begin{abstract}
We perform the first simultaneous extraction of unpolarized parton
distributions and fragmentation functions from a Monte Carlo
analysis of inclusive and semi-inclusive deep-inelastic scattering,
Drell-Yan lepton-pair production, and single-inclusive $e^+ e^-$
annihilation data.
We use data resampling techniques to thoroughly explore the
Bayesian posterior distribution of the extracted functions,
and use $k$-means clustering on the parameter samples to identify the
configurations that give the best description across all reactions.
Inclusion of the semi-inclusive data reveals a strong suppression
of the strange quark distribution at parton momentum fractions
$x \gtrsim 0.01$, in contrast with the \mbox{ATLAS} observation of
enhanced strangeness in $W^\pm$ and $Z$ production at the LHC.
Our study reveals significant correlations between the strange quark
density and the strange $\to$ kaon fragmentation function needed
to simultaneously describe semi-inclusive $K^\pm$ production data
from COMPASS and inclusive $K^\pm$ spectra in $e^+ e^-$ annihilation
from ALEPH and SLD, as well as between the strange and light antiquark
densities in the proton.
\end{abstract}

\date{\today}
\maketitle


Resolving the femtoscale structure of the nucleon remains a central
mission of ongoing and planned experimental programs at accelerator
facilities such as Jefferson Lab, RHIC, COMPASS at CERN,
\mbox{J-PARC}, and the future Electron-Ion Collider.
In particular, the flavor and spin decomposition of the proton's
valence and sea quark densities provides fascinating glimpses into
the nonperturbative QCD dynamics that give rise to the rich
phenomenology of quark and gluon interactions at long distances.
Considerable information has been accumulated from high energy
scattering on the proton's $u$- and $d$-quark parton distribution
functions (PDFs)~\cite{Blumlein13, JMO13, ForteWatt13}, and more
recently on its $\bar u$ and $\bar d$ content~\cite{Kumano:1997cy,
Vogt:2000sk, Garvey:2001yq, Melnitchouk:2017eup}.
The quantitative nature of the nonperturbative strange quark sea,
on the other hand, has remained obscured from a variety of probes
that have attempted to elucidate its structure.  This has hampered,
for example, the determination of the CKM matrix element $V_{cs}$,
as well as precision determinations of the $W$-boson mass, which
depend on precise knowledge of the strange quark
PDF~\cite{Aaboud:2017svj, Alekhin-s18}.

Since the photon couples with equal strength to $d$ and $s$ quarks,
it is difficult to disentangle the strange quark properties from
the nonstrange using purely inclusive DIS observables, even with
proton and neutron targets, without appealing to weak currents
to provide independent flavor combinations~\cite{Cooper-Sarkar15}.
The traditional method to determine the strange-quark PDF has
been through inclusive charm meson production in charged current
neutrino--nucleus DIS.
Analyses of the CCFR~\cite{CCFR95} and NuTeV~\cite{NuTeV07} $\nu$
and $\bar\nu$ cross sections from the Tevatron, and more recently
from the CHORUS~\cite{CHORUS11} and NOMAD~\cite{NOMAD13} experiments
at CERN, have yielded a strange to light-antiquark ratio
\mbox{$R_s = (s+\bar s)/(\bar u + \bar d)$} of the order $\sim 0.5$.
Unfortunately, the interpretation of the neutrino--nucleus data suffers
from uncertainties in nuclear effects in both the initial and final
states: for the former in relating nuclear structure functions to
those of free nucleons~\cite{Kalantarians:2017mkj}, and for the latter
in the treatment of charm quark energy loss and $D$ meson--nucleon
interactions during hadronization within the nucleus~\cite{Accardi09,
Majumder11}.

A method that capitalizes on the unique advantages of weak probes,
and at the same time avoids complications due to nuclear effects,
is inclusive $W^\pm$ and $Z$ boson production in $pp$ collisions.
Recent data from the ATLAS Collaboration~\cite{ATLAS_WZ12, ATLAS_WZ17}
at the LHC suggested a surprisingly larger strange quark sea than
traditionally obtained from neutrino scattering, with
$R_s \approx 1.13$ at parton momentum fraction $x = 0.023$
and scale $Q^2 = 1.9$~GeV$^2$.
Their latest analysis~\cite{ATLAS_WZ17} (``ATLAS-epWZ16'')
of the $W \to \ell \nu$ and $Z/\gamma^* \to \ell\ell$ data,
combined with the HERA runs~I and II neutral current and charged
current cross sections~\cite{HERA}, and assuming $s = \bar s$,
yielded results consistent with the earlier enhancement.

Because the ATLAS-epWZ16 fit~\cite{ATLAS_WZ17} uses only HERA and
ATLAS data, the light quark sea that emerges has $\bar d < \bar u$
at $x \sim 0.1$, in contrast to the more standard $\bar d > \bar u$
scenario found from the Fermilab E866 Drell-Yan (DY) experiment
\cite{E866, E866rat}.
In a combined fit to LHC data and charm production from neutrino DIS,
Alekhin {\it et~al.}~\cite{Alekhin-s09, Alekhin-s15, Alekhin-s18}
argued that the apparent strange quark enhancement was in fact due
to the corresponding suppression of the $\bar d$ sea at small $x$.
The ATLAS $Z \to \ell \ell$ data were found to disagree with
results from CMS~\cite{CMS_W14}, which agree with the ABMP16
global QCD analysis~\cite{ABMP16}.

The possible tension between the ATLAS and CMS measurements was
investigated in a recent dedicated analysis by Cooper-Sarkar and
Wichmann (CSKK) \cite{CSKK18}, who performed an NNLO fit to the
ATLAS and CMS inclusive $W^\pm$ and $Z$ production data at
$\sqrt{s}=7$ and 8~TeV, along with the combined HERA cross sections,
using a $K$-factor approach.
The analysis found no significant tension between the HERA, ATLAS
and CMS data, and supported an unsuppressed strange PDF at low $x$.
Their standard fit, on the other hand, gives $\bar d < \bar u$ at
$x \sim 0.1$, in contradiction with the E866 DY data, although
CSKK find that their fit with $\bar d$ forced to be greater than
$\bar u$ reduces $R_s$ only by $\approx 10\%$~\cite{CSKK18}.

From another direction, an independent source of information on the
strange-quark PDF at lower energies is semi-inclusive deep-inelastic
scattering (SIDIS), in which detection of charged pions or kaons in
the final state acts as a flavor tag of the initial state PDFs.
Earlier the \mbox{HERMES} Collaboration~\cite{HERMES08} analyzed
$K^+ + K^-$ production data from deuterons, finding a significant
rise in the extracted strange PDF at $x \lesssim 0.1$ using LO hard
coefficients, with a strong suppression at $x \gtrsim 0.1$.
A subsequent analysis~\cite{HERMES14} using new $\pi$ and $K$
multiplicity data found a less pronounced rise at small $x$,
but an essentially vanishing strangeness for $x > 0.1$.

Problems with SIDIS analyses such as that in Ref.~\cite{HERMES14},
which attempt to extract PDF information from a single data set
(in this case $K$ production) within an LO framework, were expounded
by Stolarski~\cite{Stolarski15}, who suggested additional systematic
checks of the HERMES analysis with pion production data.
Difficulties in describing the HERMES pion multiplicities were also
noted by Leader {\it et al.}~\cite{LSS14, LSS16}, who observed that
different projections of the 3-dimensional data set (which is a
function of the four-momenta of the target, $p$, virtual photon, $q$,
and produced hadron, $p_h$) do not give compatible results.

A further strong assumption in Ref.~\cite{HERMES14} and similar
analyses is that the nonstrange PDFs and fragmentation functions
(FFs) are sufficiently well known, neglecting possible correlations.
It was found in earlier analyses of polarized SIDIS data, however,
that assumptions about FFs can lead to significant differences
in extracted helicity PDFs~\cite{LSS10, LSS11}, and that a
simultaneous analysis of PDFs and FFs was needed for any definitive
conclusion~\cite{JAM16}.
Aschenauer {\it et al.}~\cite{Aschenauer15} noted that, while an NLO
analysis of semi-inclusive DIS data would be preferred, an LO extraction
is an important first step given that ``such a procedure using
semi-inclusive DIS data is not currently available.''
Later, Borsa {\it et al.}~\cite{Borsa17} considered the constraining
power of SIDIS data on the unpolarized proton PDFs through an iterative
reweighting procedure, as a further step towards a full combined
global analysis of PDFs and FFs.

In this paper, we undertake such a combined analysis at NLO,
taking advantage of recent advances in Bayesian likelihood analysis
using Monte Carlo techniques to perform the first global QCD fit
that includes SIDIS multiplicities and {\it simultaneously}
determines unpolarized PDFs and FFs.
Inclusion of the latter in the same global framework is crucial if
one is to utilize the SIDIS data without biasing the analysis
with {\it ad hoc} assumptions about FF parametrizations.
An initial attempt at a simultaneous extraction of spin-dependent PDFs
and FFs was made in Ref.~\cite{JAM17}; however, the unpolarized PDFs
there were fixed~\cite{CJ12} and spin-averaged SIDIS data were not
used in the fit.
The present work is the first of its kind to combine the
standard DIS and DY data sets used in most global fits~\cite{MMHT14,
CJ15, ABMP16, NNPDF3.1, CT14} to constrain the light-quark PDFs,
single-inclusive $e^+ e^-$ annihilation to constrain FFs,
and SIDIS multiplicities which are sensitive to both PDFs and FFs.
It thus represents an order of magnitude greater challenge than what
has ever been attempted before.


For the DIS data sets we include measurements from
  BCDMS~\cite{BCDMS},
  SLAC~\cite{SLAC},
  NMC~\cite{NMCp, NMCdop} and
  HERA runs~I and II~\cite{HERA}.
To apply the standard collinear factorization formalism~\cite{CSS}
and avoid power corrections at low energies, we include data
that satisfy the cuts on the hadronic final state mass squared
	$W^2 \equiv (p+q)^2 > 10$~GeV$^2$
and four-momentum transfer squared
	$Q^2 > m_c^2$,
where $m_c = 1.27$~GeV is the charm quark mass.
To avoid ambiguities with nuclear effects we do not consider neutrino
DIS data, and to clearly isolate the effects of the SIDIS observables
on the strange PDF we do not include the high energy LHC data in the
present analysis.

While inclusive DIS provides the mainstay observables that constrain
the PDF combinations $q^+ \equiv q + \bar q$, their ability to
isolate sea quark distributions from the valence is rather limited,
even with the presence of charged current data from HERA~\cite{HERA}.
More direct constraints on the light quark and antiquark sea
are provided by the $pp$ and $pd$ DY lepton-pair production data
from the Fermilab E866 experiment~\cite{E866}, which involves
convolutions of beam and target PDFs sensitive to small and large 
parton momentum fractions.

Further combinations of PDFs in which quark and antiquark flavors
are differentiated can be obtained from SIDIS hadron production
reactions, where a hadron $h$ is detected in the final state. 
In collinear factorization, the cross section for the inclusive
production of hadron~$h$ is given as a double convolution of the
hard scattering cross section ${\cal H}_{ij}^{\rm SIDIS}$ with the
PDF $f_i$ and parton $j \to$ hadron~$h$ fragmentation function $D_j^h$,
\begin{equation}
\frac{d\sigma_h^{\rm SIDIS}}{dx_{\rm Bj} dQ^2 dz_h}
= \sum_{ij} {\cal H}_{ij}^{\rm SIDIS} \otimes f_i \otimes D_j^h,
\label{eq.sidis}
\end{equation}
where $z_h \equiv p \cdot p_h / p \cdot q$ is the
fraction of the virtual photon's momentum carried by $h$.
Data on $\pi^\pm$~\cite{COMPASSpi} and $K^\pm$ production
\cite{COMPASSka} from COMPASS on deuterium are used for
$0.2 < z_h < 0.8$, with the low-$z_h$ cut chosen to exclude
target fragmentation and the high-$z_h$ cut avoids exclusive
channels and threshold resummation effects~\cite{Anderle:2013lka,
Accardi:2014qda}.

Lower energy SIDIS data from HERMES on hydrogen and deuterium
\cite{HERMES08, HERMES14, HERMES13} were also considered.
However, questions of compatibility of the $[x_{\rm Bj}, z_h]$ and
$[Q^2, z_h]$ projections of the data~\cite{LSS14, LSS16, Stolarski15}
as well as concerns about kinematical mass correction uncertainties
\cite{Guerrero:2017yvf} at lower $Q$ suggested that effects beyond
those included in our present framework may need to be taken into
account for a quantitative description.

While it is problematic to determine both PDFs and FFs from SIDIS
multiplicities alone, more reliable constraints on the FFs can be
obtained from hadron production in single-inclusive annihilation (SIA)
in $e^+ e^-$ collisions~\cite{JAM16, DSS07, HKNS07, AKK08, NNFF1.0,
DSSpi15, DSSk17}.
As in the previous JAM analysis~\cite{JAM16}, we consider SIA
data from
  DESY~\cite{TASSO80, TASSO83, TASSO89, ARGUS89},
  SLAC~\cite{TPC84, TPC86, TPC88, HRS87, SLD04},
  CERN~\cite{OPAL94, OPAL00, ALEPH95, DELPHI95, DELPHI98}, and
  KEK~\cite{TOPAZ95} for $Q$ up to $\sim M_Z$,
as well as more recent
results from
  Belle~\cite{Belle13, Leitgab13} 
and
  BaBar~\cite{BaBar13} 
at $Q \approx 10$~GeV.


For the QCD analysis we use hard scattering kernels computed to NLO accuracy in the $\overline{\rm MS}$ scheme, with the variable flavor number scheme for heavy flavors.
As in earlier JAM analyses~\cite{JAM16, JAM15, JAM17, JAMpi}, for the functional form
of the distributions we take the standard template, 
\begin{eqnarray}
{\rm T}(x; N, \alpha, \beta, \delta, \gamma)
&=& N x^\alpha (1-x)^\beta (1 + \gamma \sqrt{x} + \delta x),
\label{eq:T}
\end{eqnarray}
for both PDFs and FFs at the input scale, $Q^2=m_c^2$.
Typically, one template shape is needed for each nonsinglet and singlet flavor combination.
We therefore take one template function for the (nonsinglet) valence $u$- and $d$-quark PDFs, which from Regge phenomenology are expected to have a behavior $\sim x^{-1/2}$ at low $x$, and the (singlet) gluon PDF, which is expected to have the more singular $\sim x^{-1}$ behavior as $x \to 0$.
For the sea quark $\bar d$, $\bar u$, $s$ and $\bar s$ distributions, which are given by combinations of nonsinglet and singlet terms, two shapes are needed: a flavor-symmetric
sea-like shape that is dominant at low $x$, and a valence-like shape that is flavor dependent.
For the FFs one template shape was used for each $q$ and $\bar q$ flavor.

As in our previous analyses, we sample the likelihood function by
performing multiple $\chi^2$ minimizations that differ by their
initial parameters for the gradient search, as well as by the
central values of the data which are shifted via data resampling.
We use the same $\chi^2$ function as in Refs.~\cite{JAM16, JAM15,
JAM17, JAMpi}, which includes correlated systematic uncertainties
for each experiment with nuisance parameters treated on the same
footing.
For the initial analysis we fix the $\gamma$ and $\delta$ shape parameters to zero,
giving a total of 52 shape parameters together with 41 nuisance parameters for the
systematic uncertainties.
However, to explore the possible dependence of our results on the choice of
parametrization we also perform fits with $\gamma$ and $\delta$ as free parameters,
as we discuss below.

To minimize fitting bias and account for the possibility of multiple minima in the parameter space, we implement Bayesian regression using Monte Carlo methods via data resampling and a comprehensive exploration of parameter space.
To this end we devise a multi-step procedure, starting with sampling the posterior distributions for parameters using flat priors for fixed-target DIS data only~\cite{BCDMS, SLAC, NMCp, NMCdop}.
The previous step's posterior parameters then become priors for each subsequent step, in which first the DIS data sets are supplemented with the HERA run~I and II data~\cite{HERA}, followed by the DY $pp$ and $pd$ data~\cite{E866}.
At the next stage we sample the posterior distributions for the FFs using flat priors and SIA data for pions and kaons~\cite{TASSO80, TASSO83, TASSO89, ARGUS89, TPC84, TPC86, TPC88, HRS87, SLD04, OPAL94, OPAL00, ALEPH95, DELPHI95, DELPHI98, TOPAZ95, Belle13, Leitgab13, BaBar13}.
The resulting FF posteriors, together with the PDF posteriors from the previous step, are then fed in a new round where SIDIS pion and kaon data are included along with DIS, DY and SIA.

At this stage we employ a $k$-means clustering algorithm~\cite{Lloyd82, Forgy65} to identify different solutions, and use a sum of reduced $\chi^2$ values per experiment,
\begin{equation}
\overline\chi^2\ \equiv\ \sum_{\rm exp} \frac{1}{N_{\rm exp}}\chi^2_{\rm exp},
\end{equation}
where $N_{\rm exp}$ is the number of data points in each experiment, as the selection criterion.
The use of the quantity $\overline\chi^2$ ensures that the fits provide good descriptions of all data sets, not just those with the most points.
To confirm that the final solutions are a faithful representation of the likelihood function in the vicinity of the optimal parameter configuration, we construct flat priors that are confined within the posteriors identified as the best, and then perform a final run.
We stress that such an analysis would not have been feasible within a traditional approach with $\chi^2$ minimization, but has become practical within our Monte Carlo strategy.

\begin{figure}[t]
\hspace*{-0.3cm}%
\includegraphics[width=0.5\textwidth]{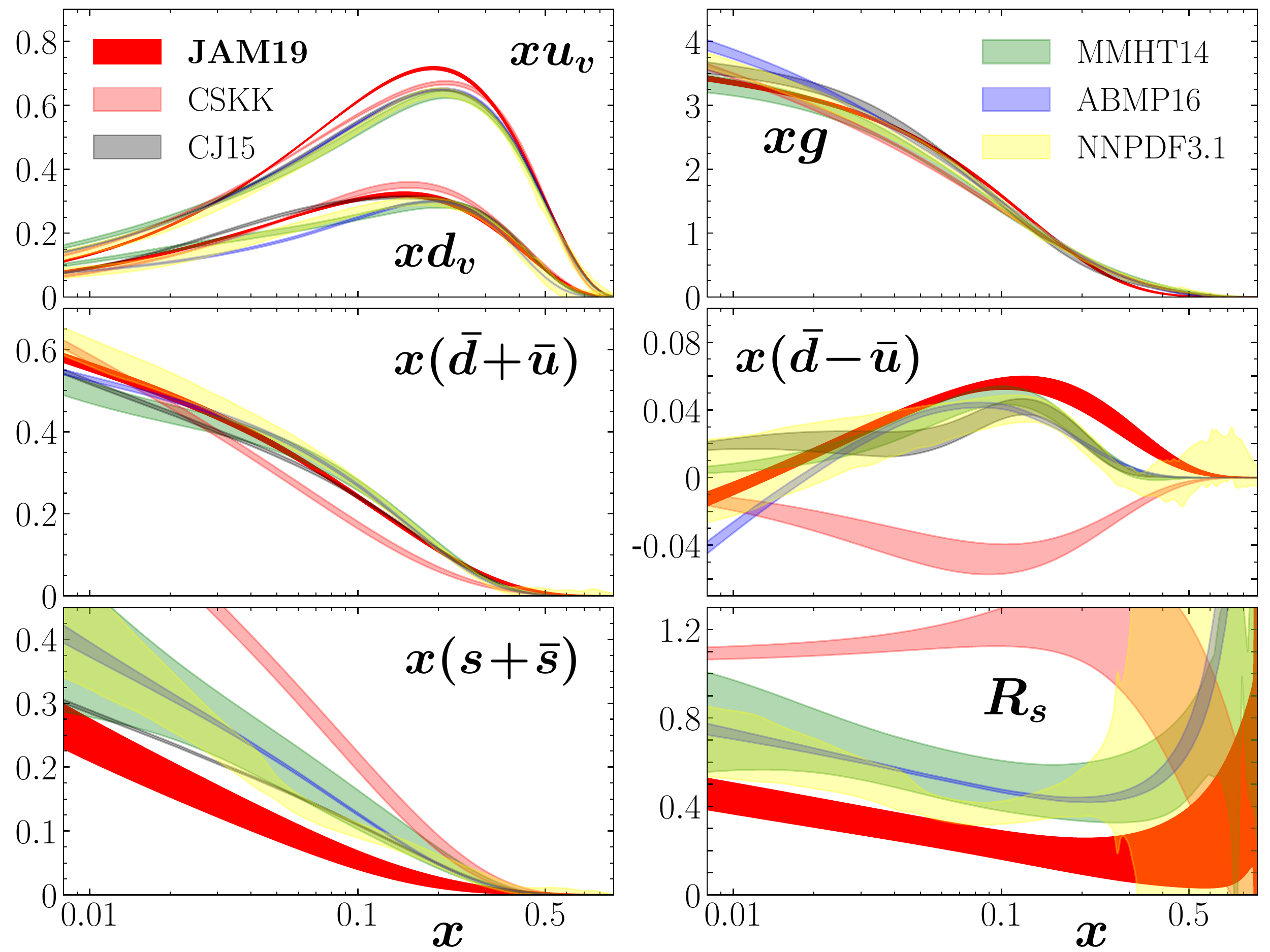}\vspace*{-0.2cm}
\caption{Comparison of the JAM19 PDFs (red bands) with the results
	from the
	CSKK~\cite{CSKK18} (pink),
	CJ15~\cite{CJ15} (gray),
	ABMP16~\cite{ABMP16} (blue),
	NNPDF3.1~\cite{NNPDF3.1} (yellow),
	and
	MMHT14~\cite{MMHT14} (green)
	parametrizations at a common scale $Q^2=4$~GeV$^2$.}
\label{fig:pdf}
\end{figure}

Our final results are based on a sample of 953 fits to 4,366 data points, giving a mean reduced $\chi^2 = 1.30$ (with individual $\chi^2$ of
	1.28 for 2,680 DIS points,
	1.25 for 992 SIDIS,
	1.67 for 250 DY, and
	1.27 for 444 SIA).
The resulting PDFs, which we refer to as ``JAM19'', are illustrated in Fig.~\ref{fig:pdf} at a common scale of $Q^2=4$~GeV$^2$.
Our results for the nonstrange distributions are generally similar to those obtained by other groups~\cite{MMHT14, CJ15, ABMP16, NNPDF3.1, CT14}.
Some differences appear in the valence $u$-quark distribution in the region $0.05 \lesssim x \lesssim 0.2$, where the JAM19 result sits slightly about the others.
As discussed below, this appears correlated with the strong suppression of the strange-quark PDF found in our combined analysis of the PDFs and FFs.
We note that in the electromagnetic DIS structure functions, which provide the bulk of the constraints on the PDFs in this region, the quark and antiquark distributions enter additively, so that a suppression of the strange PDF will be compensated by a slight enhancement in the valence distributions.

To test the dependence of the valence-quark PDFs on our chosen parametric form, we have also performed a Monte Carlo fit where the polynomial parameters $\gamma$ and $\delta$ in Eq.~(\ref{eq:T}) were allowed to vary.
The results indicate that the changes are rather small with the more flexible parametrization, on the scale of the uncertainties, and suggest that our valence PDFs do not depend significantly on whether $\gamma$ and $\delta$ are free parameters or are set to zero.
To further explore the flexibility beyond the one-shape scenario with nonzero $\gamma$ and $\delta$, we also performed fits with two basic template shapes from Eq.~(\ref{eq:T}), with $\gamma=\delta=0$,
        $N_1 x^{\alpha_1}(1-x)^{\beta_1}$ 
      + $N_2 x^{\alpha_2}(1-x)^{\beta_2}$ 
for both the $u_v$ and $d_v$ distributions.
Again, the results were almost indistinguishable from those of our default JAM19 analysis shown in Fig.~\ref{fig:pdf}.

For the $\bar d - \bar u$ asymmetry, significant differences exist between our results and the CSKK fit~\cite{CSKK18}, which uses only HERA and LHC results and excludes the fixed-target DY data~\cite{E866, E866rat}.
The latter force a positive asymmetry peaking at $x \gtrsim 0.1$, in contrast to the negative $\bar d - \bar u$ driven by the HERA data.
For the gluon distribution at low~$x$ the main constraint is from the HERA data.

\begin{figure}[bt]
\hspace*{-0.1cm}%
\includegraphics[width=0.49\textwidth]{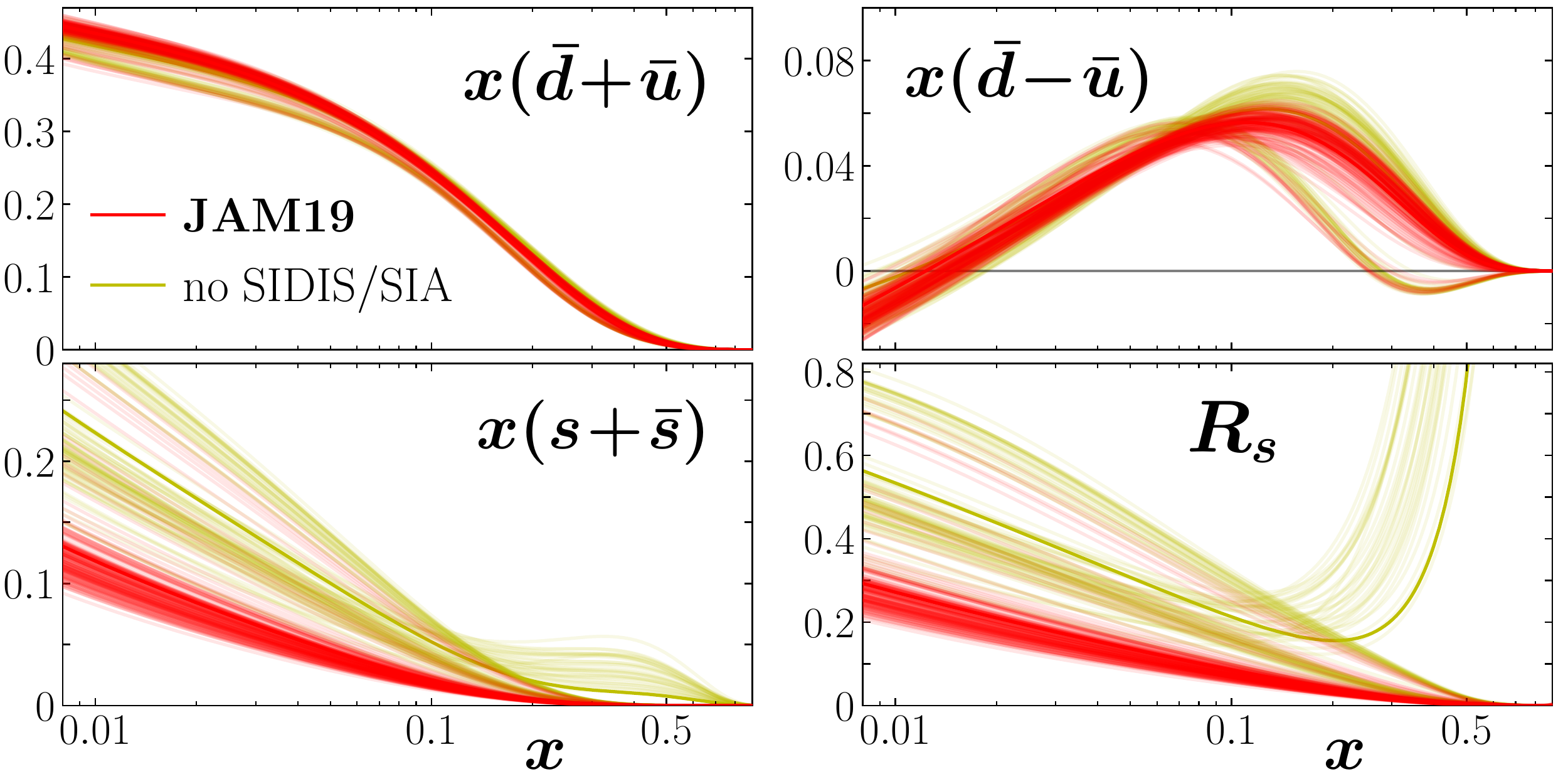}\vspace*{-0.2cm}
\caption{Comparison of the light and strange sea-quark PDFs in the
	JAM19 analysis (red lines) with fits excluding SIDIS and SIA
	data (yellow lines) at the input scale.}
\label{fig:sidis_pdf}
\end{figure}

The most striking result of our analysis is that the strange-quark PDF is significantly
reduced compared with that reported by ATLAS~\cite{ATLAS_WZ12, ATLAS_WZ17} and the CSKK
fit~\cite{CSKK18}.
For the strange to nonstrange ratio, we find
	$R_s \approx 0.2-0.3$
at $x \sim 0.02$, in contrast to values of $R_s \sim 1$ inferred from the ATLAS data,
and closer to those extracted from neutrino experiments.
(We note that inclusion of the neutrino--nucleus DIS data, as used by a number of the
global QCD analysis groups, would enhance the strange-quark signal up to $R_s \sim 0.5$,
thereby effectively requiring smaller values for the nonstrange or valence quark PDFs
to describe the data --- see Fig.~\ref{fig:pdf}.)
The most significant source of the strange suppression is the SIDIS and SIA $K$
production data, as Fig.~\ref{fig:sidis_pdf} illustrates.
Without these data, the $s^+$ PDF is poorly constrained, in contrast to the light flavor
sea, which is not strongly affected by the SIDIS multiplicities.
Consequently, while the ratio $R_s$ varies over a large range without SIDIS (and SIA)
data, and at low $x$ is compatible with $R_s \sim 1$, once those data are included its
spread becomes dramatically reduced.

\begin{figure}[t]
\hspace*{-0.3cm}
\includegraphics[width=0.49\textwidth]{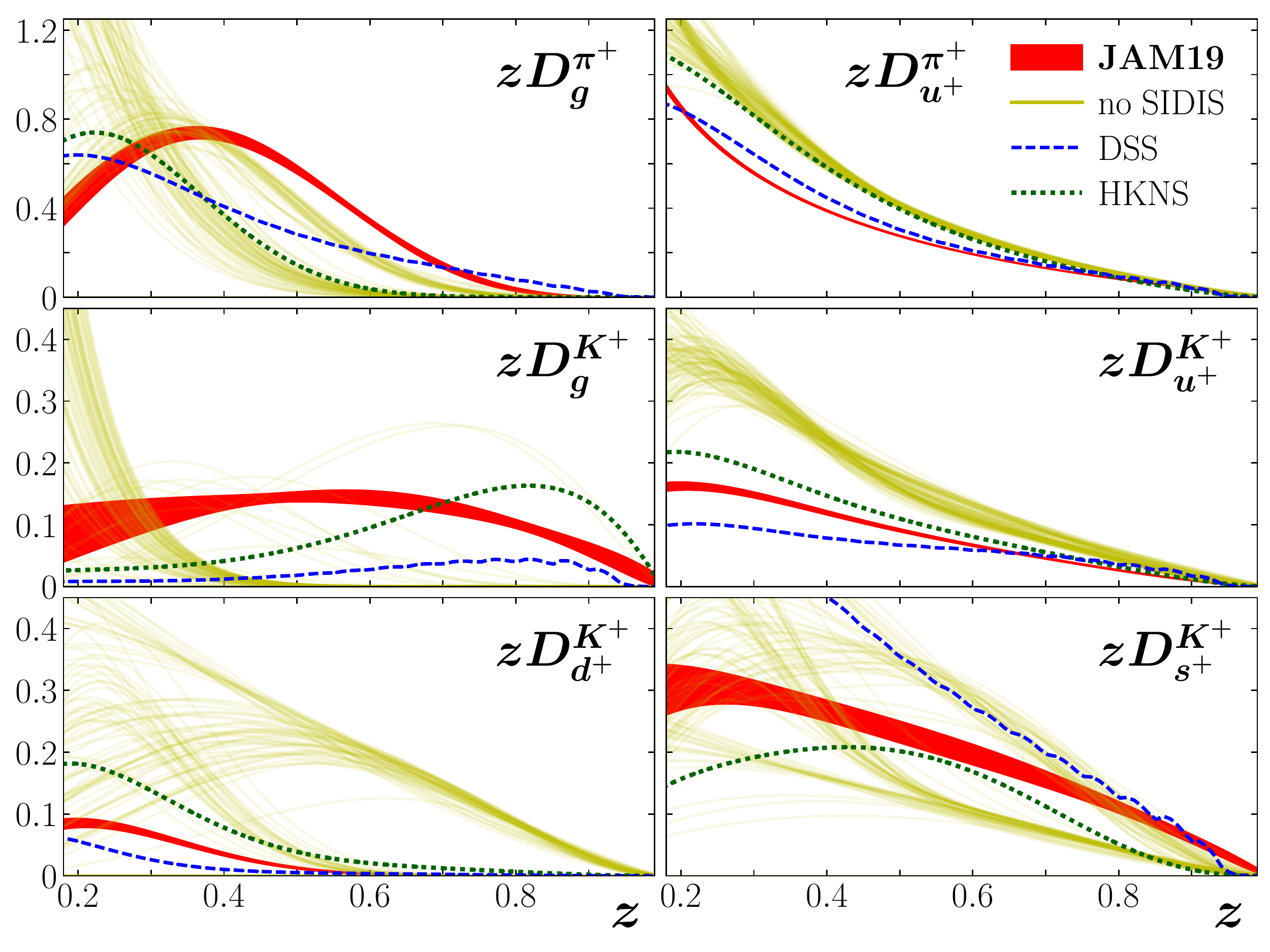}\vspace*{-0.2cm}
\caption{Comparison of the JAM19 FFs (red bands) into $\pi^+$
	(for $g$ and $u^+$) and $K^+$ mesons (for $g$, $u^+$, $d^+$
	and $s^+$) with results of fits without SIDIS data
	(yellow lines) and from the
	DSS~\cite{DSS07} (blue dashed) and
	HKNS~\cite{HKNS07} (green dotted)
	parametrizations at the input scale.}
\label{fig:ff}
\end{figure}

The vital role played by the SIDIS and SIA measurements can be better appreciated from the FFs, shown in Fig.~\ref{fig:ff}, where we compare the full results with those constrained only by SIA data, and with some common FF parametrizations~\cite{DSS07, HKNS07}.
Since the SIA data alone cannot discriminate between $q$ and $\bar q$ fragmentation, we show the FFs for $q^+ \to \pi^+, K^+$.
While the pion FFs are generally in better agreement, the kaon FFs display more variation.
The $\chi^2$ per datum values from our full fit are 1.07 for the $\pi^\pm$ SIA data and 1.48 for the $K^\pm$ SIA data.
For the SIDIS data we find $\chi^2$/datum of 1.18 for pions and 1.30 for kaons.
These values are generally comparable to those found by other groups~\cite{DSS07, HKNS07, DSSpi15, DSSk17}, although in some cases different data are fitted and none of the other analyses performs a simultaneous analysis as we do here.

Our full fits reflect the standard hierarchy of the favored and unfavored fragmentation, with the $s^+ \to K^+$ FF larger than the $u^+ \to K^+$, which in turn is larger than the unfavored $d^+ \to K^+$.
In contrast, for the SIA-only fits the unfavored $D_{d^+}^{K^+}$ includes solutions with both soft and hard shapes, the latter being correlated with a small $\bar s \to D^{K^+}$ fragmentation.
In Ref.~\cite{JAM16} the FFs were constrained also by light-flavor tagged SIA data, which are not included here because of potential bias from their reliance on Monte Carlo simulations.
Instead, we find that the combination of SIDIS and SIA data forces the favored $D_{\bar s}^{K^+}$ to be large at high $z$, comparable to the DSS fit~\cite{DSS07}.

The large $D_{\bar s}^{K^+}$ ($= D_s^{K^-}$) found in our combined analysis has major consequences for the strange-quark PDF.
Since the $K^+$ SIDIS deuterium cross section is given by the
flavor combination
	$2 (u+d) D_u^{K^+} +\, \bar s D_{\bar s}^{K^+}$,
at moderate $x$ and $z \gg 0$ it is dominated by the $u$-quark term.
The $K^-$ cross section, in contrast, is proportional to the combination
	$2 (\bar u+\bar d) D_{\bar u}^{K^-} +\, s D_s^{K^-}$,
and receives comparable contributions from strange and nonstrange quarks.
Because the nonstrange PDFs are much better determined, and the nonstrange favored
    $D_u^{K^+} = D_{\bar u}^{K^-}$
is well constrained by the SIA and SIDIS data (see Fig.~\ref{fig:ff}), the $K^+$ and $K^-$ SIDIS multiplicities provide sensitivity to the total strange quark contribution, $s\, D_s^{K^-}$.

\begin{figure}[t]
\vspace*{-0.8cm}%
\hspace*{-0.7cm}%
\includegraphics[width=0.55\textwidth]{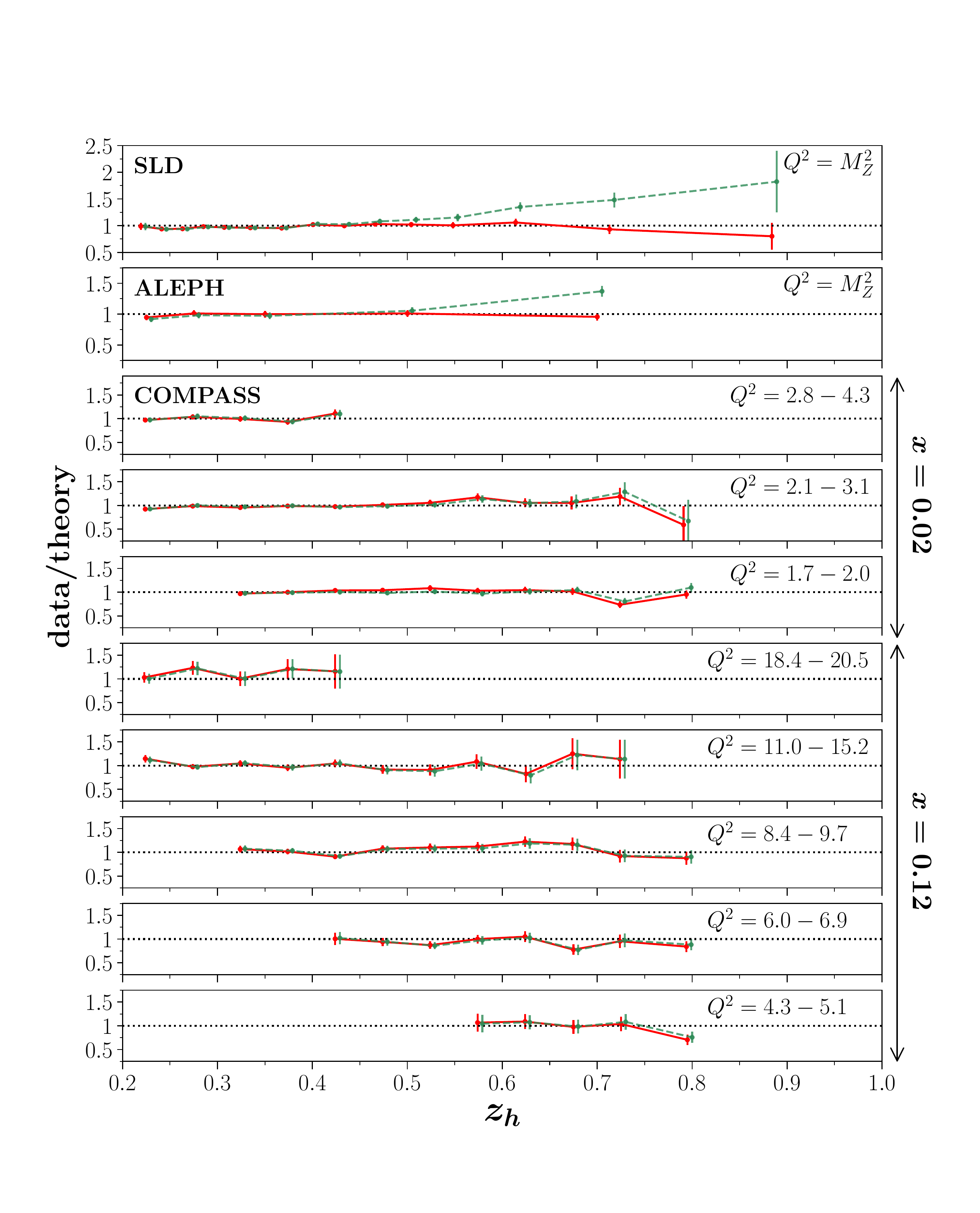}\vspace*{-0.9cm}
\caption{Ratios of data to theory for $e^+ e^- \to K^\pm X$ cross sections
versus $z_h$ from  SLD~\cite{SLD04} and ALEPH~\cite{ALEPH95} (top two panels),
and for $K^-$ production in SIDIS from COMPASS~\cite{COMPASSka} for two
representative $x_{\rm Bj}$ bins ($x_{\rm Bj}=0.02$ and 0.12) for the $Q^2$
ranges indicated (in GeV$^2$). The red points correspond to the average
of the best solutions selected by the $k$-means algorithm, while the
green points represent unfavored solutions with smaller $D_{s^+}^{K^\pm}(z)$
and larger $s(x)$.}
\label{fig:sidisK-}
\end{figure}

In practice, the SIDIS data alone admit solutions which have {\it either} a relatively small $s(x)$ and large $D_s^{K^-}\!(z)$, {\it or} a large $s(x)$ and small $D_s^{K^-}\!(z)$, as the data/theory ratios in Fig.~\ref{fig:sidisK-} illustrate.
The solutions with the best $\chi^2$ after $k$-means clustering are illustrated by the red points in Fig.~\ref{fig:sidisK-}, and correspond to the full results (with the small $s$-quark PDF and large $D_s^{K^-}$ FF) displayed in Figs.~\ref{fig:pdf}--\ref{fig:ff}.
The green points in Fig.~\ref{fig:sidisK-} represent solutions that give equally good descriptions of SIDIS data, but with a large $s(x)$ weighted by a small $D_s^{K^-}\!(z)$, which then underestimates the SIA cross sections by $\sim 50\% - 100\%$ for large $z_h$ values.
For example, for the SLD~\cite{SLD04} and ALEPH~\cite{ALEPH95} data illustrated in Fig.~\ref{fig:sidisK-}, the best solutions (red) yield an average reduced $\chi^2_{\rm SLD} = 1.38$ and $\chi^2_{\rm ALEPH} = 0.74$, but much larger $\chi^2$ values (4.10 and 4.62, respectively) are found for the unfavored solutions (green).
Such correlations are symptomatic of the inverse problem for nucleon structure, and our analysis clearly indicates that the path towards its solution must involve simultaneous extraction of all collinear distributions within a single unified framework.

%
The SIDIS $K^\pm$ production data could also in principle discriminate
between the $s$ and $\bar s$ PDFs, which need not have the same $x$
dependence~\cite{Signal87, Malheiro97, Thomas00, Catani04, Wang16,
Sufian18}.
We explored this scenario by parametrizing $s(x)$ and $\bar s(x)$
separately, but found that none of the SIDIS or other data sets
showed clear preference for any significant $s-\bar s$ asymmetry
within uncertainties.
Future high-precision SIDIS data from Jefferson Lab or from the
planned Electron-Ion Collider may allow more stringent determinations
of the $s$ and $\bar s$ PDFs~\cite{Aschenauer19}, as would inclusion
of $W\, +$ charm production data from the LHC~\cite{ATLAS_Wc14,
CMS_Wc14}, with better knowledge of $c$-quark jet fragmentation
and hadronization~\cite{CSKK18}.

In addition to collinear distributions, SIDIS data will also provide opportunities in future to study transverse momentum dependent PDFs and FFs, which involve more complicated correlations between the longitudinal and transverse momenta and spins of partons~\cite{TMDs}.
The methodology developed here for the simultaneous global QCD analysis of different types of distributions will pave the way towards universal analyses of quantum probability distributions that will map out the 3-dimensional structure of the nucleon~\cite{EIC}.

\section*{Acknowledgements}

We thank A.~Accardi, J.~Qiu and T.~Rogers for helpful discussions and communications.
This work was supported by the US Department of Energy contract DE-AC05-06OR23177, under which Jefferson Science Associates, LLC operates Jefferson Lab. J.E. was partially supported by the Netherlands Organization for Scientific Research (NWO).


\end{document}